\documentclass[11pt, a4paper]{article}

\usepackage[export]{adjustbox}
\usepackage[skip=5pt]{caption}
\usepackage[all]{nowidow}
\usepackage{comment}
\usepackage{amsmath,amsfonts,amssymb}
\usepackage{graphicx}
\usepackage[colorlinks=true, allcolors=blue]{hyperref}
\usepackage[numbers]{natbib}
\usepackage{doi}
\usepackage{authblk}
\usepackage[font={small},labelfont=bf]{caption}
\usepackage{arxiv}

\title{Astrophotonic Technologies}

\author[a]{Aline N. Dinkelaker}
\affil[a]{Leibniz Institute for Astrophysics (AIP), An der Sternwarte 16, 14482 Potsdam, Germany}

\begin{document} 
\maketitle

\begin{abstract}
Over the past two decades, photonics have been developed as technological solutions for astronomical instrumentation for, e.g., near-infrared spectroscopy and long baseline interferometry. With increasing instrument capabilities, large quantities of high precision optical components are required to guide, manipulate, and analyze the light from astronomical sources. Photonic integrated circuits (PICs) and fiber-based devices offer enormous potential for astronomical instrumentation, as they can reduce the amount of bulky free-space optics and pave the way for innovative solutions. Astrophotonic devices are particularly interesting for interferometry due to their compact design on the centimeter scale, even for a large number of telescope inputs. Already, astrophotonic components are integrated in high-end instruments at the VLTI and at the CHARA Array. Photonic beam combiners at wavelengths from visible to mid-infrared have been fabricated using lithographic techniques or ultrafast-laser inscription, with several components tested on-sky. This paper will provide a glimpse into the growing field of astrophotonics, its current status and potential for new technologies.  
\end{abstract}

\section{INTRODUCTION}
\label{sec:intro}  

Advances in astronomy depend on different factors going hand in hand, including the construction of additional or larger telescopes, improvement of instrument functionality and precision, as well as appropriate data analysis and theoretical simulation with sufficient computing power. In the context of astronomical instruments, in particular for near-infrared (NIR) spectroscopy and long baseline interferometry, photonic components have been developed as technological contributions to pushing the limits and capabilities of modern instruments. As these photonic components are tailored and applied to astronomy, they are known as astrophotonics. Their development has been an active field of research ever since its first consideration more than 20 years ago, see Refs.~\citenum{Bland-Hawthorn:09, Bryant:17, Dinkelaker:21} for collections, Refs.~\citenum{Bland-Hawthorn:17,Norris:2019,Gatkine2019State,Minardi:2021,Labadie:2022, Roth2023} for overviews, a recently published book on the \textit{Principles of Astrophotonics}~\cite{Ellis:2023:book}, and the \textit{2023 Astrophotonics Roadmap}~\cite{Jovanovic:2023} outlining current status, challenges and potentials of this technology. 

\begin{figure}
    \centering
    \includegraphics[width=0.9\linewidth]{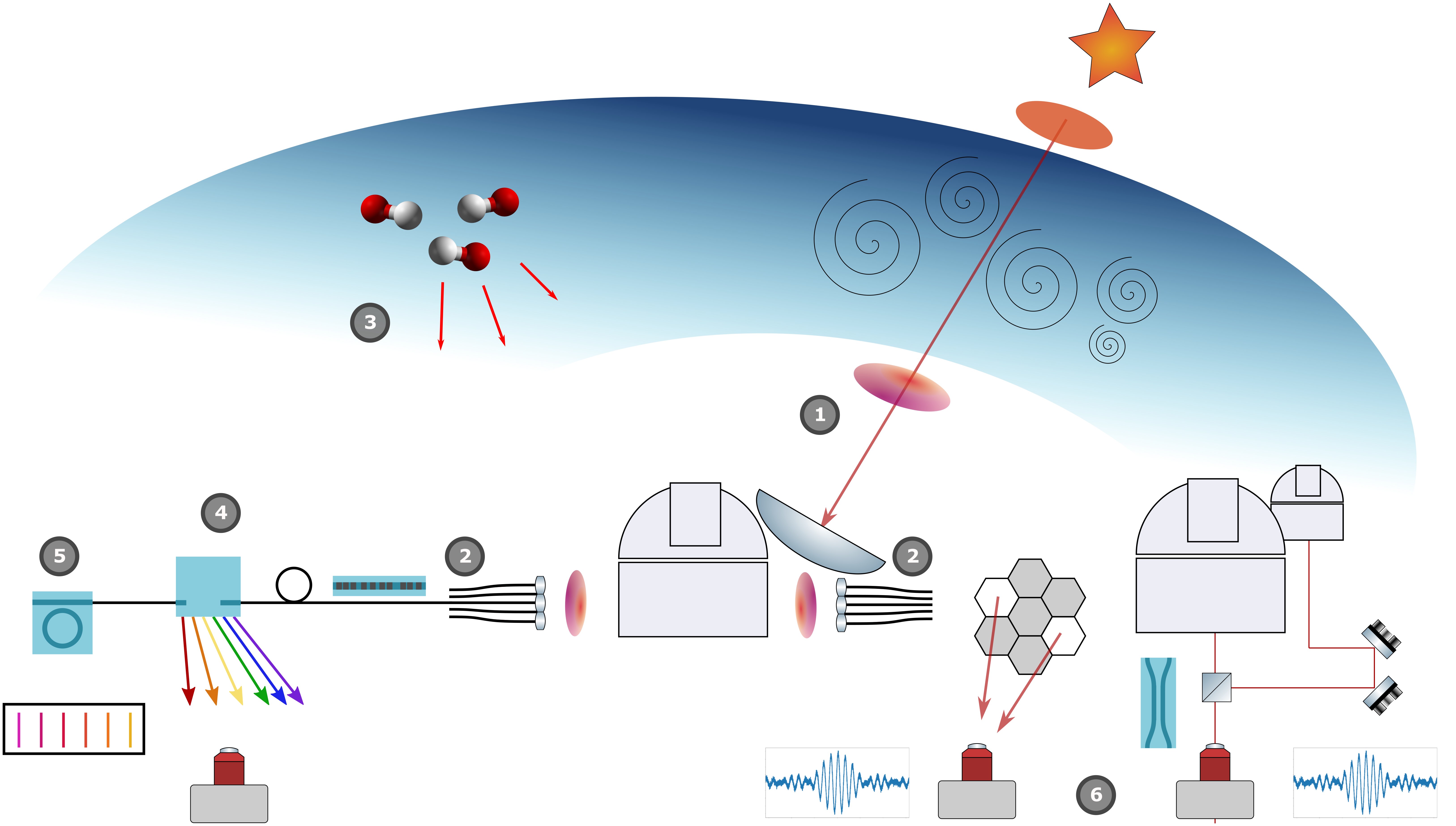}
    \caption{Schematic (simplified and not to scale) of possible beam path and component chains. Astrophotonics can be used to fulfill different functions along the beam path after the telescope: from filtering unwanted light from the atmosphere to interferometric beam combination. The round labels with the numbers refer to individual applications, for which astrophotonic components are presented as examples (see text for details): 1) atmospheric distortion and photonic lanterns, 2) photonic reformatters, 3) filters for atmospheric emissions, 4) spectrographs, 5) frequency references, 6) interferometric beam combiners. }
    \label{fig:astrobeampath}
\end{figure}

Photonic components provide structures in optical fibers or chips to guide and manipulate light, through different refractive indices, materials, and geometries. Here, designs, components, expertise, and fabrication from the telecommunication industry are an important foundation for the development of astrophotonics~\cite{Roth2023}, making components for NIR wavelengths particularly accessible and efficient, e.g. optical fibers for the astronomical H-band (centered around 1650~nm) that overlaps with telecommunication wavelengths around 1550~nm, which are commercially available, affordable and low-loss. While this is a good starting point, astrophotonic research continuously expands the available wavelength range and functionality of astrophotonic devices, stretching from the visible to the mid-infrared (MIR). Astrophotonic components can be used at different points along the beam path and in astronomical instruments, see Fig.~\ref{fig:astrobeampath} for a schematic. 

This paper aims to provide a short, general overview of the astrophotonic landscape and different aspects that have to be considered in this interdisciplinary field, and to highlight some of the technologies and components as examples. 

\subsection{Photonic Building Blocks for Astronomy}
\begin{figure}[t]
    \centering
    \includegraphics[width=0.45\linewidth]{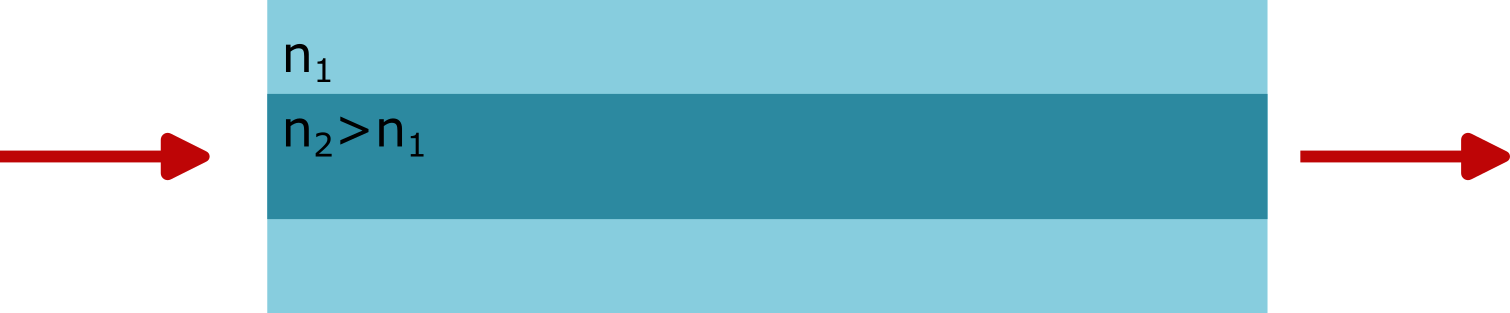}
    \caption{Simplified schematic of a WG structure: two regions in a material have different refractive indices, $n_1$ and $n_2$, where the darker region has a refractive index $n_2 > n_1$ in order to guide light. The red arrows indicate the light coming in from the left and exiting the WG on the right. }
    \label{fig:Schem-WG}
\end{figure}

\begin{figure}[t]
    \centering
    \includegraphics[width=0.3\linewidth, valign=c]{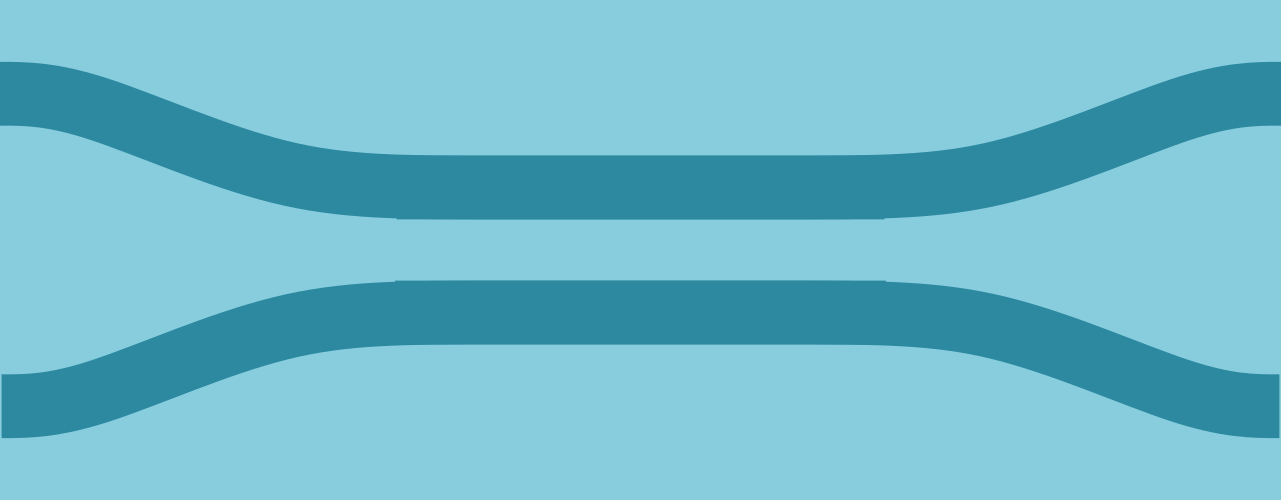}
    \hspace{8pt}
    \includegraphics[width=0.2\linewidth, valign=c]{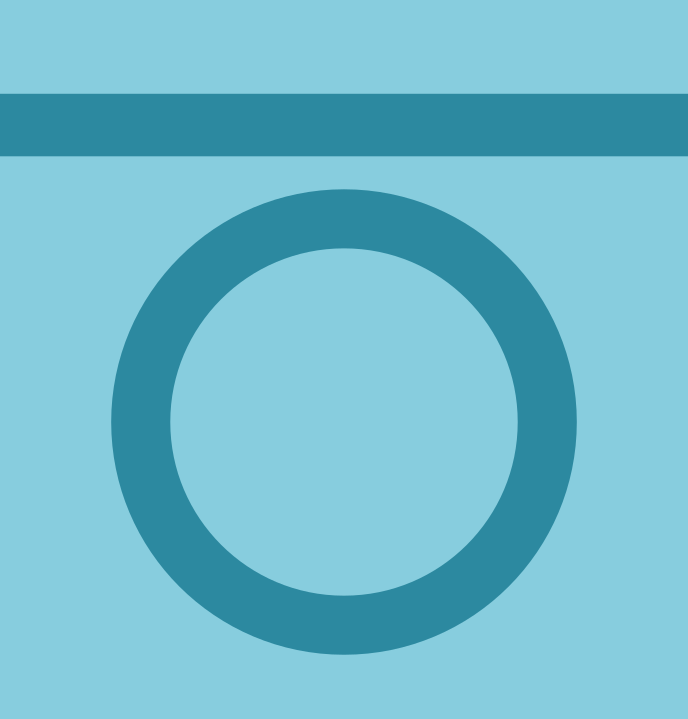}
     \hspace{8pt}
    \includegraphics[width=0.35\linewidth, valign=c]{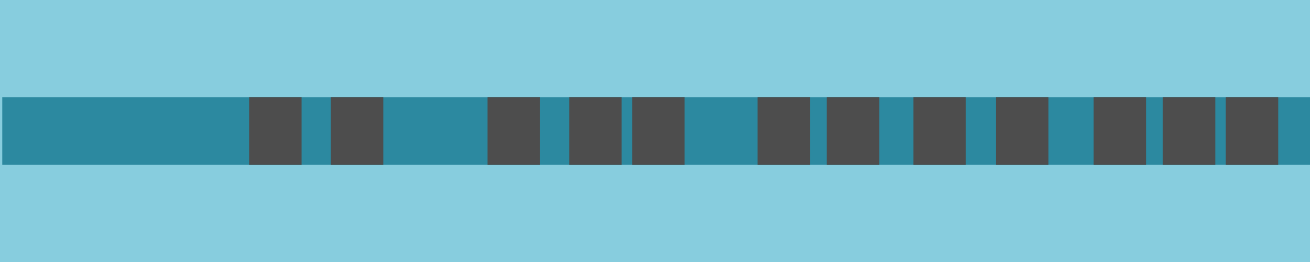}
    \caption{Schematic representation of different possible WG structures, such as curves that bring two WGs close to each other, which can be used to make directional couplers (left), a ring structure (center), or grating structures made by changing the refractive index inside a WG (right). These (simplified) structures are some of the fundamental geometries for photonic components. }
    \label{fig:Schem-structures}
\end{figure}

Generally, there are photonic solutions for various functionalities along the beam path from the telescope up to the detector, see Fig.~\ref{fig:astrobeampath}. The underlying principle is the guidance of light in a geometric structure of material that is surrounded by material with lower refractive index, a waveguide (WG), see Fig.~\ref{fig:Schem-WG} (more details can be found in e.g. Ref.~\citenum{Ellis:2023:book}). The unique properties of photonics can be exploited for applications in astronomy: 
\begin{itemize}
    \item evanescent effects due to the small geometries, where light couples from one WG to other WGs that are close enough ($\mu$m-range), see Fig.~\ref{fig:Schem-structures},
    \item spatial filtering, as photonic WGs are typically single-mode (SM),
    \item ability for mode conversion (specifically the photonic lantern, PL),
    \item confining the light in WGs allows for curved structures, which is particularly useful for remapping - and not possible using bulk optics under normal circumstances. 
\end{itemize}

In addition to providing functionalities that might otherwise be difficult to achieve with free space optics, many of the advantages of photonics lie in their form factor:

\begin{itemize}
    \item Photonic devices typically have footprints on the order of cm, which can reduce size, weight, and cost of instruments.
    \item Monolithic photonics have the potential to be less sensitive to environmental effects (vibrational, thermal), which will improve instrument stability.
    \item With the right interface design (e.g. fiber connection), modular upgrades and extensions are possible. 
\end{itemize}

Different geometries, such as those shown (simplified) in Fig.~\ref{fig:Schem-structures} can be combined to achieve the desired functionality of a component. This includes components with WG geometries for light distribution, remapping, spatial filtering and modal control, lenslets for coupling, structures for wavelength filtering (e.g. fiber bragg gratings, FBG, or chains of ring resonators) as well as for dispersion (e.g. arrayed waveguide gratings, AWG), wavelength referencing (e.g. frequency combs, FC), and interferometric beam combination. These components can replace individual components or whole subsystems of the traditional bulk optics that are usually found in astronomical instruments. (Astro-) Photonic research ranges from fundamental physics at the level of light-matter interaction to delivery of scientific data in commissioned instruments at telescopes. 

\subsection{Astrophotonics for the Next Generation of Telescopes and Arrays}

Upgrades to existing instruments, future space-based telescopes, as well as astronomical instruments for the next generation of telescopes, in particular ESO's Extremely Large Telescope (ELT)~\cite{ELT}, require innovative technological solutions if instrument size, weight, and cost are to be limited. 

A larger telescope diameter will enhance the number of collected photons as well as the diffraction limited resolution. However, larger telescope diameters also mean that atmospheric effects will be harder to correct, even more so for light at shorter wavelengths, which puts an emphasis on (extreme) adaptive optics (AO) systems. Even after (partial) correction using an AO system, atmospheric distortions result in a larger beam diameter, requiring larger (bulk) optics, e.g. a massive grating in the spectrograph part of an instrument, in order to analyze all of the light. Additional astrophotonic solutions such as the photonic lantern (PL)~\cite{Birks:15} can aid to facilitate the use of light with imperfectly corrected wavefront~\cite{MacLachlan:2016,Diab:2020}, and astrophotonic spectrographs can provide space-saving alternatives at each output of a PL~\cite{Madhav:2024}. Thus, future telescopes surpassing 10~m in diameter will see substantial technology development in its wake, including astrophotonics. 

\begin{figure}
    \centering
    \includegraphics[height=110pt]{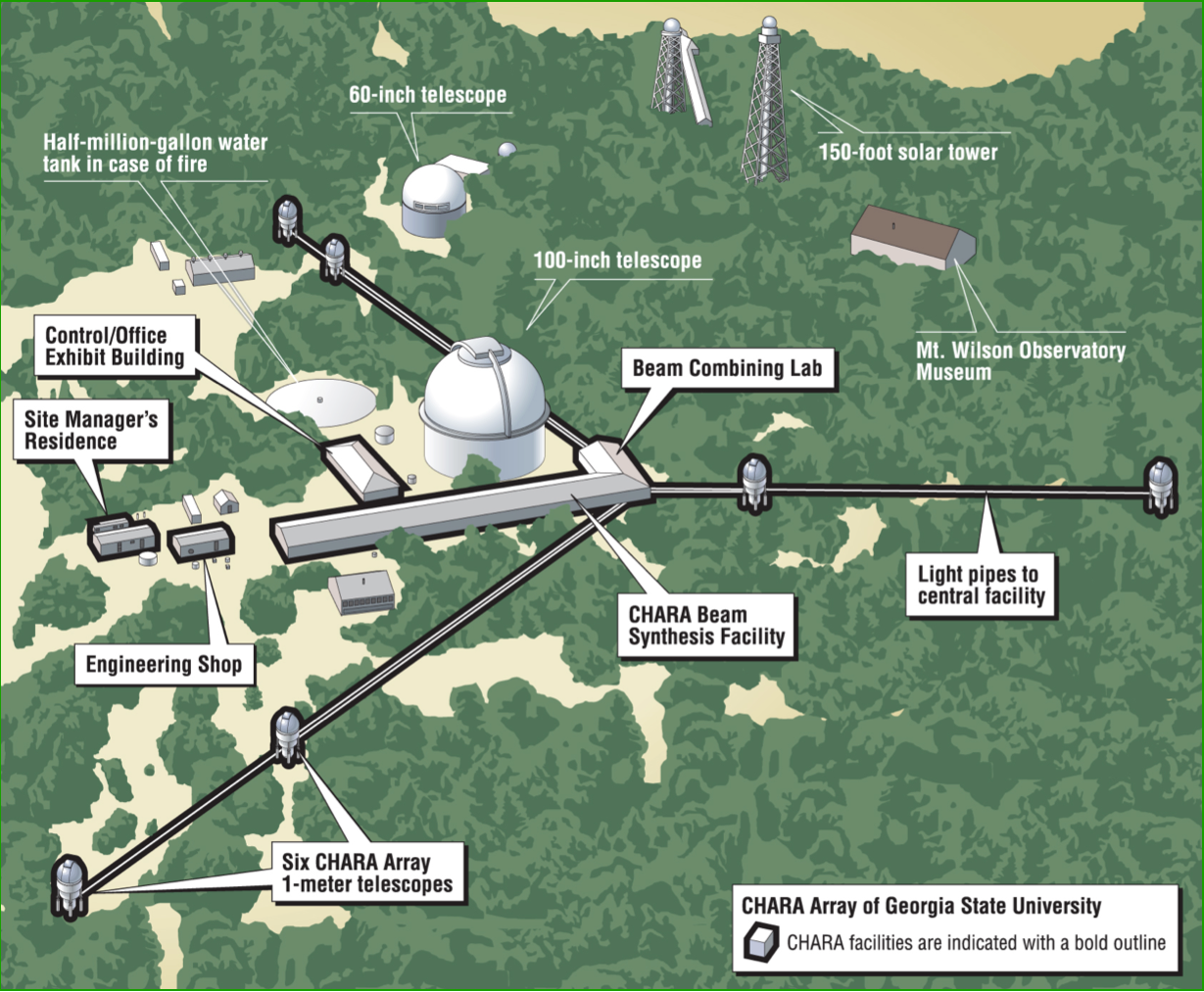}
    \includegraphics[height=110pt]{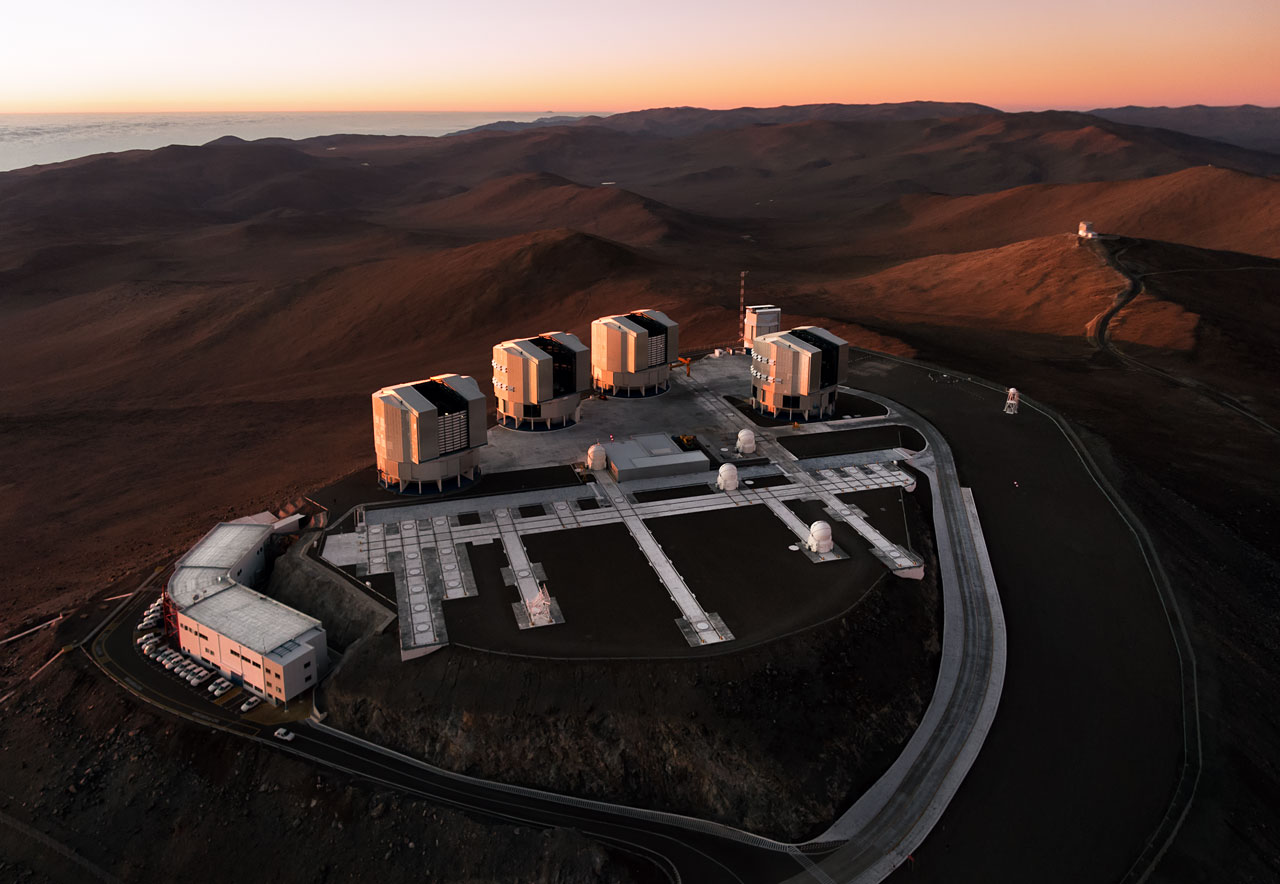}
    \includegraphics[height=110pt]{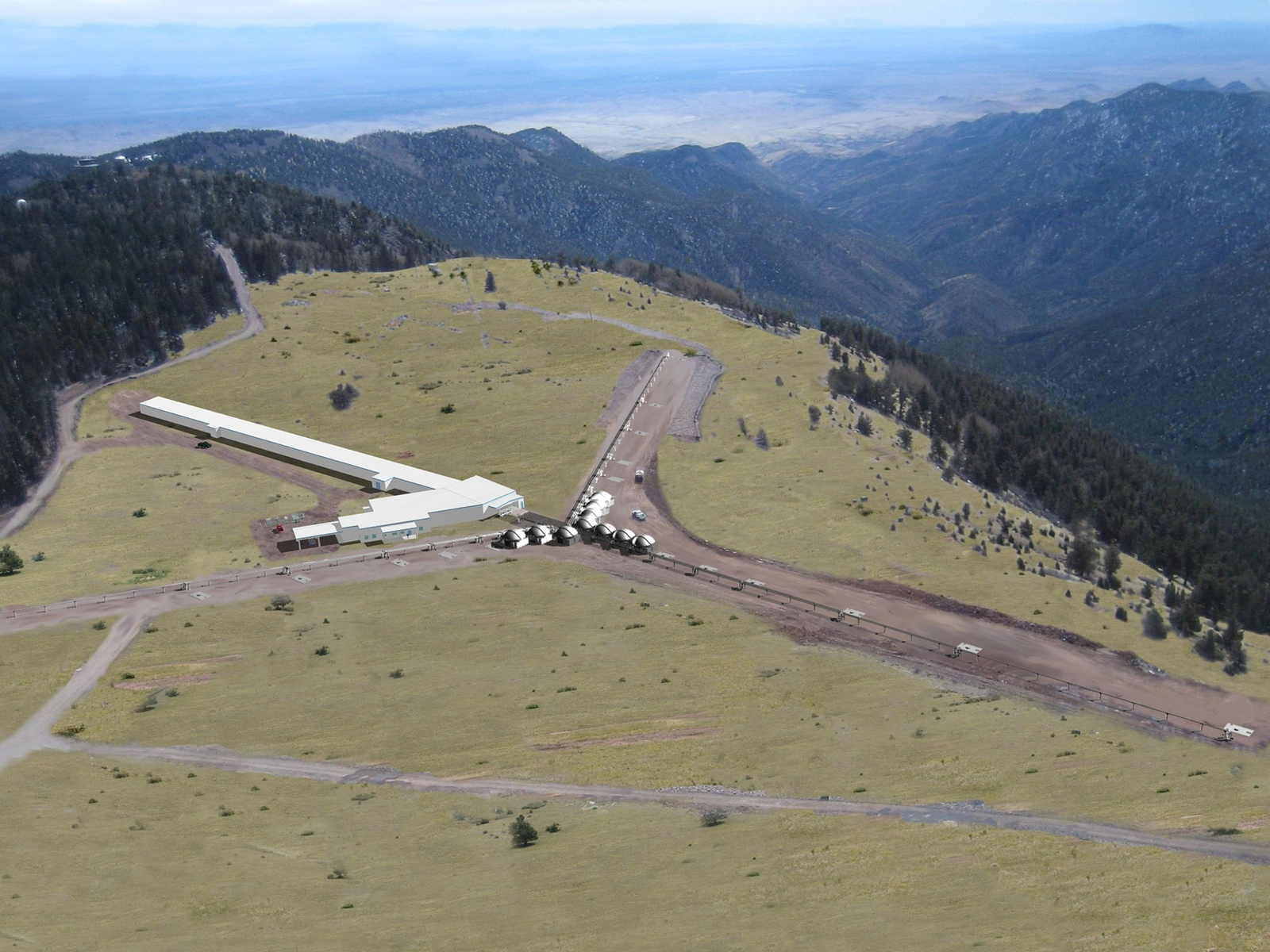}
    \caption{Interferometric telescope arrays. (Left:) The layout of the CHARA Array on Mount Wilson in California, with its six 1-m telescopes and the buildings that house the delay lines and the beam combiner instruments. Also visible are the other telescopes on Mount Wilson, which are not part of CHARA. The planned 7th telescope is not yet included in this figure. Image downloaded from the CHARA website~\cite{CHARA}, credit: The Observatories of the Carnegie Institution. (Center:) A picture of ESO's Paranal Observatory, with the four 8.2-m Unit Telescopes (UTs) as well as the four 1.8-m Auxiliary Telescopes (ATs) and their set of rails for the VLTI~\cite{VLTI}. The 4.1-m Visible and Infrared Survey Telescope for Astronomy (VISTA) can be seen in the distance. Image credit: G.~Hüdepohl (atacamaphoto.com)/ESO. (Right:) Plans for the MROI in New Mexico foresee ten 1.4-m telescopes, the second Unit Telescope has already arrived. Image downloaded from the MRO website~\cite{MRO}, credit: MRO.}
    \label{fig:arrays}
\end{figure}

Similarly, the extension of long baseline interferometric arrays to larger number of telescopes increases the amount of required beam combinations, where the number of telescopes, $N$, leads to $\frac{N\cdot(N-1)}{2}$ beam combinations. Currently, the Very Large Telescope Interferometer (VLTI)~\cite{VLTI} and the Center for High Angular Resolution Astronomy (CHARA) Array~\cite{CHARA, Brummelaar:2005} provide the most telescopes in an array and the longest baselines for long baseline interferometry in the optical/NIR, with four (VLTI) and six (CHARA) telescopes, see Fig.~\ref{fig:arrays}. This is scheduled to change: after constructions are complete, the Magdalena Ridge Observatory Interferometer (MROI) will consist of ten telescopes~\cite{Creech2019,MRO}, see Fig.~\ref{fig:arrays}. In addition, integration of the seventh (mobile) telescope at CHARA has already started~\cite{Ligon:2024,Scott:2024}. This will provide opportunities for photonic beam combiners to showcase their capabilities, as they enable light combination from multiple telescopes inside devices with footprints on the centimeter scale. Already, long baseline interferometry engages photonics, with GRAVITY at VLTI very successfully employing its beam combiner chip~\cite{Benisty:2009}, and MIRC-X/MYSTIC at CHARA using fibers for image-plane beam combination of all six beams~\cite{Anugu:2020}.  

Their physical dimensions and stability make astrophotonics a potentially favorable option for space-based instruments. While no large baseline interferometer has been realized in space yet, current research towards the space-based MIR nulling interferometer LIFE (Large Interferometer For Exoplanets)~\cite{LIFE, Quanz:2022, Glauser:2024} considers if and how astrophotonic beam combiners can replace free space optics.

\section{DEVELOPMENT AND MANUFACTURING}
The development of photonic components constitutes of different phases, which each or all might be iterated multiple times: 
\begin{enumerate}
    \item simulation and modeling, resulting in a design,
    \item manufacturing, using e.g. lithographic methods, laser writing, 3D printing or other ways of glass processing,
    \item laboratory characterization on a testbench in a (reasonably) controlled environment using a laser and/or broadband light source for optical characterization of the photonic component, 
    \item on-sky testing, which provides the necessary step to achieve maturity in a realistic telescope environment, to verify the functionality and identify potential shortcomings for further device optimization. 
\end{enumerate}
Even after successful on-sky operation, a photonic device might be subject to redesign, optimization, or additional integration to form more complex photonic integrated circuits (PIC) and deliver enhanced functionality. 

\subsection{Examples of Fabrication Methods}
\label{fab}
\begin{figure}[t]
    \centering
    \includegraphics[height=170pt, valign=t]{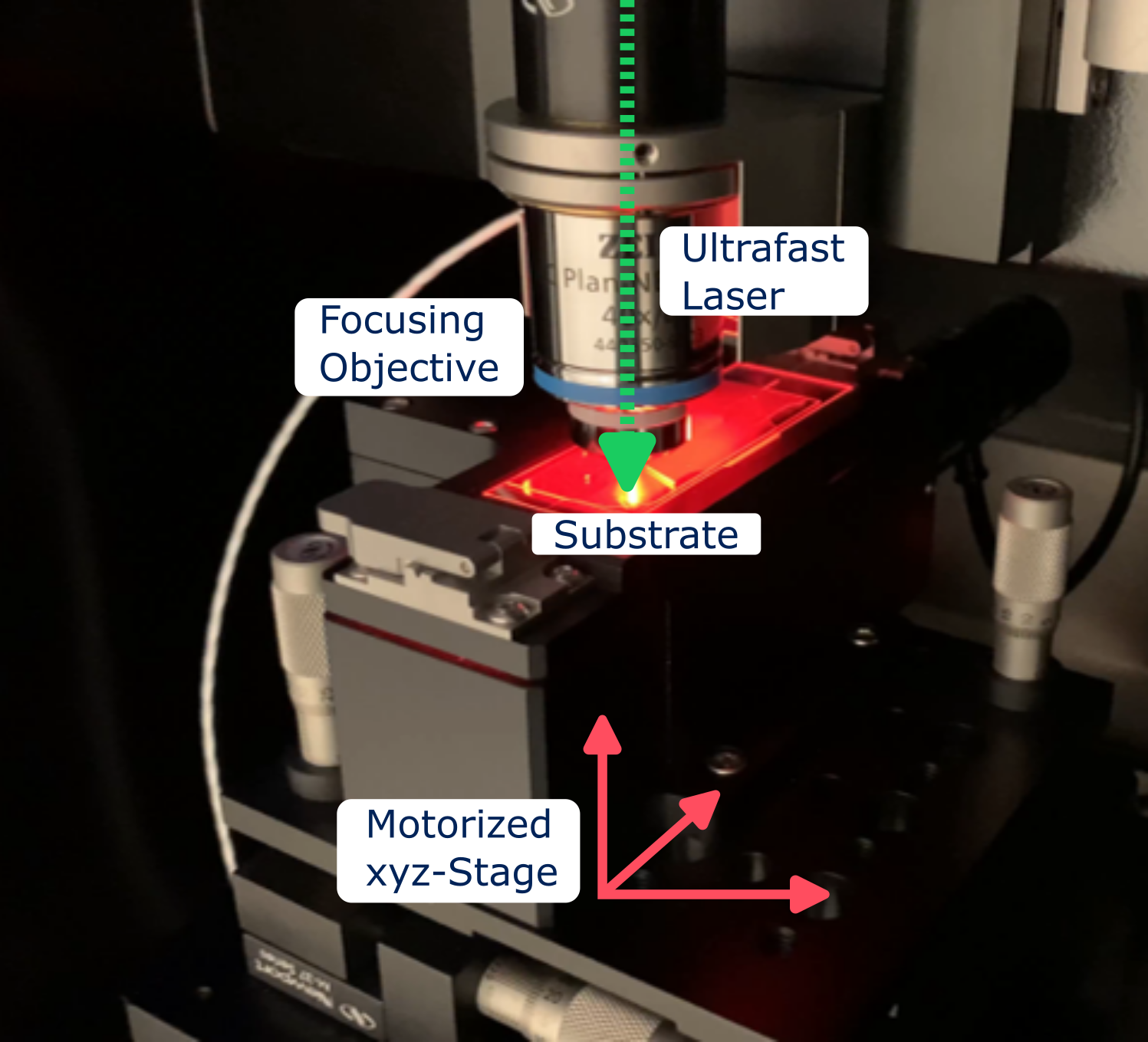}
    \hspace{8pt}
    \includegraphics[height=170pt, valign=t]{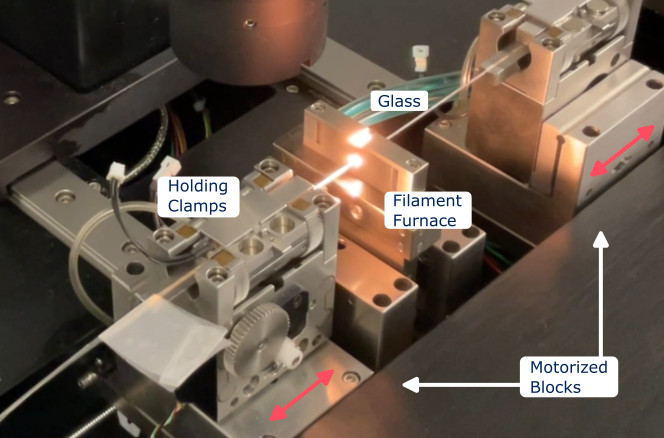}
    \caption{Two examples of fabrication setups for astrophotonics: (Left:) a setup for ULI (here: MKS FemtoFBG, image credit: A.~V.~Mayer (AIP)), showing the objective that focuses the femtosecond laser onto a substrate, which is mounted on an automated translation stage. (Right:) A glass processing and tapering setup (here: Vytran glass processor, image credit: J.~Rypalla (AIP), see also Ref.~\citenum{Rypalla:2024}), where a fiber bundle (or multi-core fiber) is stacked inside a glass capillary and mounted on two motorized blocks. When the filament heats the glass to a melting point, the blocks create a predefined pull, thus tapering the glass to the designed geometry. }
    \label{fig:Manu}
\end{figure}

As the dimensions of the WG structures that have to be fabricated are on the order of the wavelength of light (at least the WG width is typically several $\mu$m), the manufacturing process and its properties and tolerances as well as materials and their variations contribute to device performance. This includes losses due to imperfections and WG roughness, precision in splitting ratios, and predictability and repeatability of manufacturing a device. There are different platforms and fabrication methods, each with their own advantages. They will not be described in any detail here, but to give an idea of the topic, an incomplete list of frequently used methods is included (see also Refs.~\citenum{Labadie:2016, Jovanovic:2023, Ellis:2023:book} for more information or Refs.~\citenum{Gatkine:2021, Thomson:09, Davenport:2021, Anagnos:2021} for examples of each of the  techniques):
\begin{itemize}
    \item \textbf{Lithography (or more precisely: photolithography)} is a multi-step process, where the WGs and geometry of a complete device are first designed on a mask. The design is then transferred from the mask onto the wafer, i.e. the material for the chip (for PICs, a wafer typically consists of two layers). The structures are created using a sequence of special coatings and light exposure in combination with etching for material removal. The designs are generally 2D. (See e.g. Ref.~\citenum{Gatkine:2021}.)
    \item \textbf{Direct laser writing} / \textbf{ultrafast-laser inscription (ULI)} uses an ultrafast-laser (typical pulse duration of a few hundred fs) to locally change the refractive index of a (glass) substrate at the focus point of the laser through non-linear and thermal processes. By moving the substrate relative to the laser focus, 3D WG structures can be created, with WG diameters (for NIR light) on the order of 10~$\mu$m, which are typically several cm long. (See Fig.~\ref{fig:Manu} for an example of such a setup and Refs.~\citenum{Thomson:09, Piacentini:2022} for more information.) The changes induced in the substrate material by an ultrafast-laser also enable the combination of this fabrication technique with selective chemical etching (Ultrafast-Laser Assisted Etching, ULAE), which can be used to fabricate e.g. microlenses~\cite{Ross:2020} or image-slicers~\cite{Gorp:2024}.   
    \item \textbf{Glass heating and tapering} is typically used to transform the waveguiding properties of fibers and fiber bundles. The glass is heated locally and the fiber(s) pulled with changing velocity, thereby creating tension to shape and taper. (See Fig.~\ref{fig:Manu} and Ref.~\citenum{Davenport:2021} for examples of a setup and fabrication, respectively.)
    \item \textbf{3D printing} / \textbf{additive manufacturing} can be used to create micro-optics, even directly on the facet of a chip or fiber, e.g. micro-lenses to couple light into a WG. (See e.g. Ref.~\citenum{Anagnos:2021}.)
\end{itemize}

For all manufacturing methods, there are several aspects to be optimized:
\begin{itemize}
    \item reduce optical losses in a device,
    \item more deterministic fabrication, i.e. being able to predict the outcome using models,
    \item more repeatable fabrication, i.e. creating multiple devices with identical properties and performance.
\end{itemize}
For the fabrication of photonic chips, lithography and ULI differ in several aspects, including the available materials and the refractive index difference that can be achieved, but also with respect to the manufacturing approach: with lithography, all design choices have to be made at the start of a production run in form of the mask, and manufacturing is usually done externally, which can come with high cost and a long waiting time. Thus, design changes and developments are harder to try out. The results, however, can be very precise, with high confinement and low loss. ULI on the other hand allows for rapid prototyping and a high level of flexibility - provided that access to a ULI system is possible. To initially find suitable writing parameters is not easy, since the parameter space is incredibly large: parameters include laser power, focus spot size, and writing speed - all of which also depend on the type of laser (typically either Ti:Sapphire or ytterbium doped solid state lasers) and its pulse duration, as well as the substrate material and other specifics (see also Ref.~\citenum{Stoian:Nano} for fundamental light-matter interactions and different methods using ultrafast-lasers). Once a range of parameters is found, this technique enables a wealth of possibilities for 3D WG structures. 

For all of the fabrication methods listed above, the choice of material depends on the wavelength of light that is to be analyzed. The choice of fabrication method depends on the application and the approach that fits best. Most devices can be fabricated using different methods. The following section will present several examples of astrophotonic devices, but there are many more components with other functionalities, as well as similar components that have been fabricated using alternate methods. 

\section{ASTROPHOTONIC COMPONENTS AND DEVICES}

Different astrophotonic components can be used at various points along the beam path to guide and manipulate light, see Fig.~\ref{fig:astrobeampath}. This section will provide some general examples of astrophotonic components before presenting examples with specifically interferometric applications in Sec.~\ref{astrophotonicsforinterferometry}. For images of the components, please refer to the references. The components presented could also be combined to form more integrated photonics-based devices and instruments, see e.g. Ref.~\citenum{Norris:2019}. 

\begin{figure}[t]
    \centering
    \includegraphics[height=100 pt]{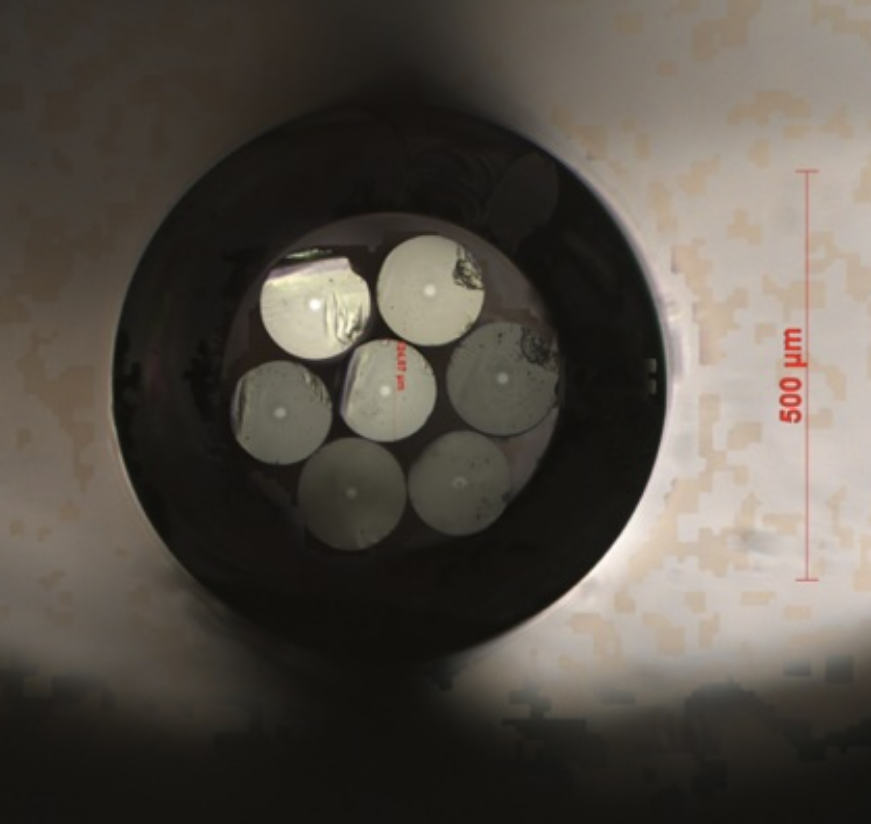}
        \includegraphics[height=100 pt]{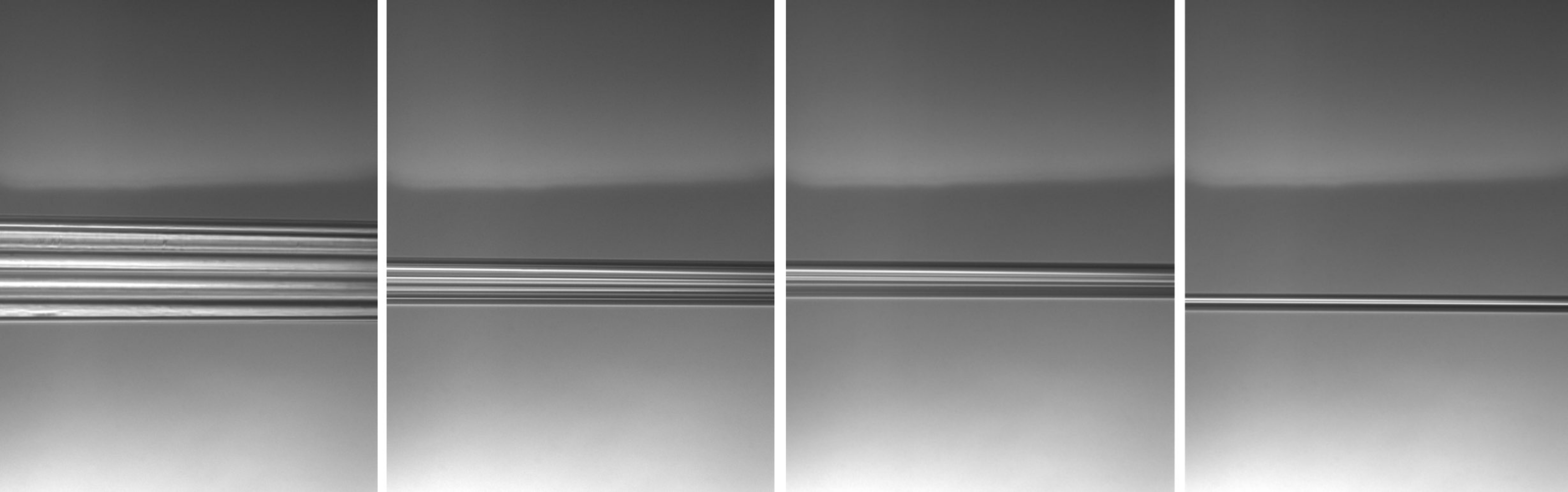}
    \caption{Examples of a PL at different stages of fabrication from a stack of SM fibers. (Left:) Cross-section microscope image of a fiber bundle inside a capillary at the start of tapering. (Right:) Sideview microscope images at different positions along the taper. Images from J.~Rypalla (AIP)~\cite{Rypalla:2024}.}
    \label{fig:PL}
\end{figure}

\paragraph{Photonic lanterns:} When light from an astronomical source passes through turbulent atmosphere, the wavefront will become distorted instead of being a plane wave (see Fig.~\ref{fig:astrobeampath}, label 1). Without adaptive optics (AO) that can correct the wavefronts at least partially, coupling into SM waveguides -as is required for most astrophotonic components- will be very inefficient. Photonic lanterns (PL)~\cite{Leon-Saval:2005} can improve the fraction of light that can be used in instruments by converting the multi-mode (MM) light (number of modes $N_M$) from the uncorrected / partially corrected wavefront to several SM outputs ($N_{PL}$), see e.g. Ref.~\citenum{Leon-Saval:2005, Diab:2020}. For efficient mode transfer in a PL, the number of output modes must be (at least) the number of input modes, i.e. $N_{PL} = N_M$, and the transition between the two regions must sufficiently gradual (adiabatic). Each of the SM output WGs can then be used to include additional photonic components, which would ideally all be identical to each other. Depending on the type of PL, reformatting functionalities can also be combined with PLs to redistribute the light to a more suitable configuration (see also label 2 in Fig.~\ref{fig:astrobeampath}), e.g. a pseudo-slit can be formed using photonic chip~\cite{MacLachlan:2016} or fiber~\cite{Yerolatsitis:2017} reformatters to reduce the spectrograph size. PLs can be fabricated in different ways: 
\begin{itemize}
\item using the 3D capabilities of ULI to write SM waveguides that converge to a MM waveguide~\cite{Thomson:2011},
\item adiabatically tapering down a multi-core fiber (MCF) to produce a MM waveguide at one end~\cite{Jovanovic:2020},
\item adiabatically tapering down a stack of SM fibers inside a capillary to produce a MM waveguide at one end (see also Refs.~\citenum{Davenport:2021, Rypalla:2024} and Fig.~\ref{fig:PL}).
\end{itemize}

While PLs can be used in either direction (MM-input to SM-output or vice versa), the application described here requires light to enter at the MM end to be distributed over the SM ports. The number of PL ports has to match the number of modes at the input for efficient conversion. Practical limitations come from e.g. the fabrication methods or available resources such as fibers and capillaries and their required refractive indices and geometries (core/cladding for fibers and inner/outer diameter for capillaries).

\paragraph{OH suppression filters:}

One of the components that can be combined with a PL are hydroxyl (OH) suppression filters, which could be added to each SM output of the PL. Especially in the NIR wavelength range, observations on ground are disturbed by a large number of bright but narrow OH emission lines originating from the atmosphere (illustrated in Fig.~\ref{fig:astrobeampath}, label 3), with time-varying intensity, see Ref.~\citenum{Ellis:2008} for more details. Ideally, these would be filtered out to increase the signal-to-noise ratio of ground-based observations, but the requirements are difficult to be met: high (30 - 40 dB) suppression ratio, high throughput for the science light between the emission lines, narrow linewidths, many ($> 100$) filter notches. Existing classical solutions (e.g. dispersion of light and masking the unwanted emission line wavelengths) suffer from scattered light~\cite{Ellis:2008, Rahman:2020}. Thus, photonic components are under development, which show promising results. To approach these requirements and fabricate a filter that rejects a large number of specific wavelengths, a complex component has to be fabricated with high precision. Two candidates are:
\begin{itemize}
    \item Fiber Bragg Gratings (FBGs), where an aperiodic grating (via changes of the refractive index) is inscribed into a fiber core (See Fig.~\ref{fig:Schem-structures} for an illustration). The refractive index change for the grating can be achieved in different ways. One method to generate the FBG OH suppression filters is to design a phase mask with which a grating pattern is imprinted onto a fiber using an utraviolet (UV) laser~\cite{Rahman:2020}. This method allows for very repeatable grating fabrication, which is important to fabricate identical components to be used in combination with PLs. The design of the phase mask to create the desired filter is a complex process; a current phase mask development is presented in Ref.~\citenum{Luo:2024}. 
    \item A sequence of ring resonators of different sizes (on the ~10~$\mu$m scale), each designed to filter a specific wavelength, is lithographically fabricated within the same photonic chip~\cite{Ellis:17,Kuhn:2024}. 
\end{itemize}
FBGs have already been developed and tested with the GNOSIS~\cite{Trinh:2013} and PRAXIS~\cite{Content:2014,Ellis:2020} instruments, suppressing $> 100$ emission lines, and efforts to improve performance, fabrication methods, stabilization~\cite{Alvarez:2024} and integration of OH suppression filters are ongoing. 

\paragraph{Dispersion elements / spectrograph on a chip:}
With SM fibers, a chip-based spectrograph element can be used to disperse the source light into its spectral components on a detector for analysis (see Fig.~\ref{fig:astrobeampath}, label 4 for illustration). Typical bulk optic dispersion elements include prisms and Volume Phase Holographic Gratings (VPHGs). One particular example of an astrophotonic dispersive element is the Arrayed Waveguide Grating (AWG). 
\begin{figure}[t]
    \centering
    \includegraphics[width=0.6\linewidth]{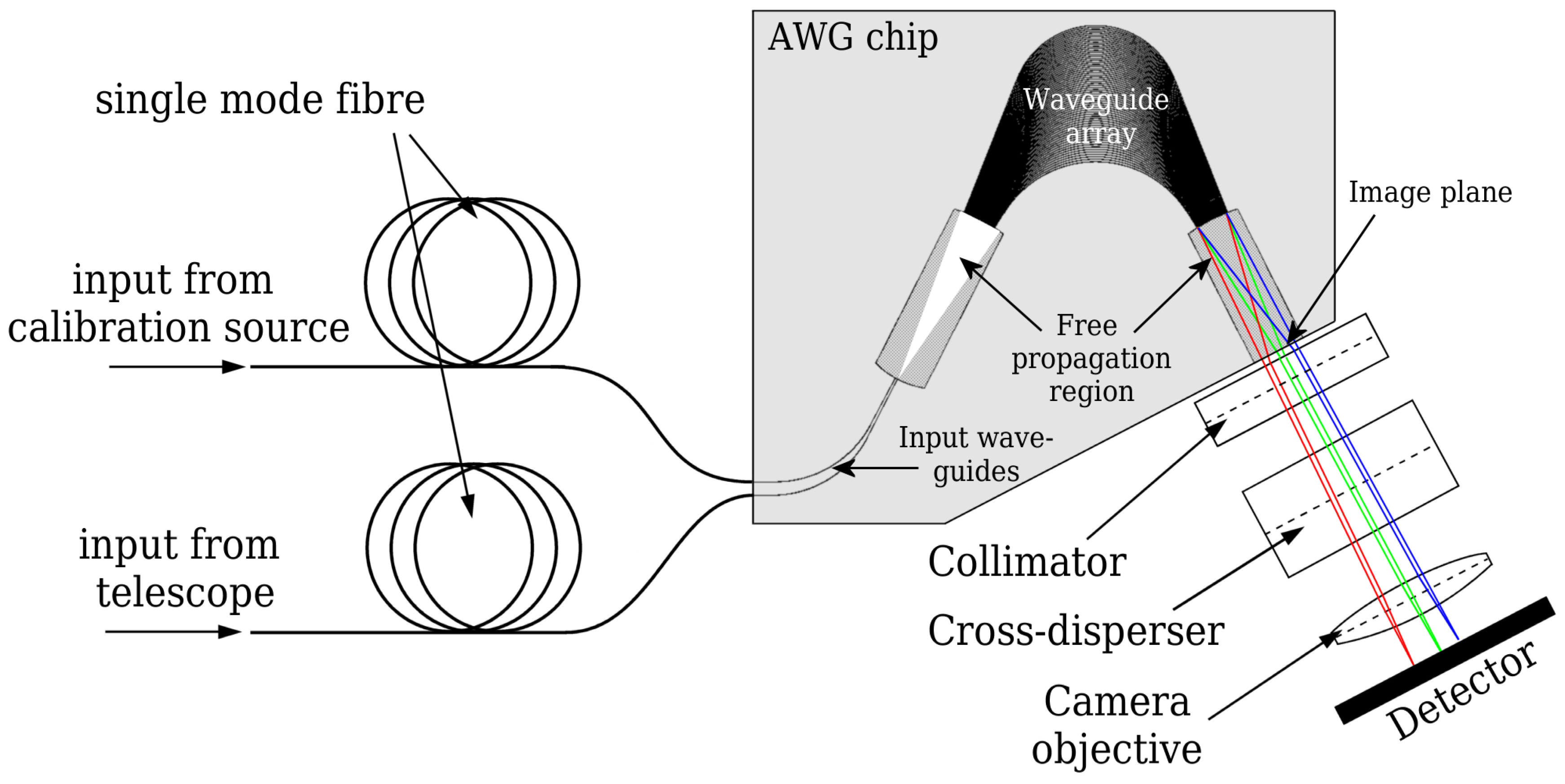}
    \caption{Schematic of an AWG (image from Ref.~\citenum{Stoll:2017}), indicating the input of the science light as well as a reference light on the left, followed by the first free propagation region, the waveguide array, and the second free propagation region at the output. There are additional free-space optics for collimation and cross-dispersion, before the light is distributed onto a detector.} 
    \label{fig:AWG}
\end{figure}
An AWG (see Fig.~\ref{fig:AWG} for a schematic~\cite{Stoll:2017}) consists of input WGs and a free propagation region at the start, followed by the main part: the waveguide array. Here, hundreds of WGs are arranged such that the length increases for each WG (e.g. using a curved configuration, in which the outer WGs are longer than the inner ones). The light exits the array into a second free propagation region at the output of the chip. Starting from this point, free-space optics are mainly used to collimate and cross-disperse the light (to avoid overlap of different orders), and to bring it onto the detector. Such AWGs can achieve high resolving powers $R = \frac{\lambda}{\Delta \lambda}$ (e.g. R~$\approx10000 - 28000$ in the astronomical H-band~\cite{Stoll:2021b}). Besides the resolving power, important performance metrics are transmission and broadband operation, and a trade-off can be made according to the application's requirement~\cite{Gatkine:2024OE}. For an overview and comparison of different AWGs, see the tables and details in Ref.~\citenum{Madhav:2023}, and see Refs.~\citenum{Roth2023, Stoll:2017} and references within for more details. 

Integration of AWG chips into instrument has already begun, e.g. a chip as described above~\cite{Stoll:2021b} in the PAWS instrument~\cite{Roth2023, Madhav:2023, Hernandez:2024}, with space-based instruments being potential applications for such small components after sufficient verification. Their size also allows for combination of components: a cascaded design of AWG chips~\cite{Gatkine:2024} can enable a larger bandwidth for broadband operation. Stacks of AWGs can accept light from multiple PLs, as AWGs can simultaneously disperse light from several fibers~\cite{Madhav:2024}. Ongoing work aims to exploit this for the development of a Multi-Object Spectrograph (MOS) incorporating MM fibers feeding into PLs, which are coupled to a stack of AWGs~\cite{Madhav:2024}. 

\paragraph{Frequency references:}
In addition to the spectrally dispersed science light, a reference spectrum (Fig.~\ref{fig:astrobeampath}, label~5) is required to identify the absolute positions or any changes in the spectral lines. Classically, spectral line lamps (or absorption cells) are used to obtain a reference, but these have several disadvantages, such as widely and unevenly spaced emission lines and uneven intensities. To measure relative shifts of the science light wavelength, evenly spaced and stable reference lines are required. This is particularly important for exoplanet detection using the radial-velocity measurements, where small Doppler shifts in the emission lines of a star indicate its motion due to an orbiting exoplanet. To achieve the desired Doppler accuracy for exoplanet detection of different types requires the measurement of radial velocities on the order of m/s (even down to several cm/s). Thus, reference spectra need to be stable on the hundreds of kHz level --and better-- over long timescales. 

\begin{figure}[t]
    \centering
    \includegraphics[width=0.7\linewidth]{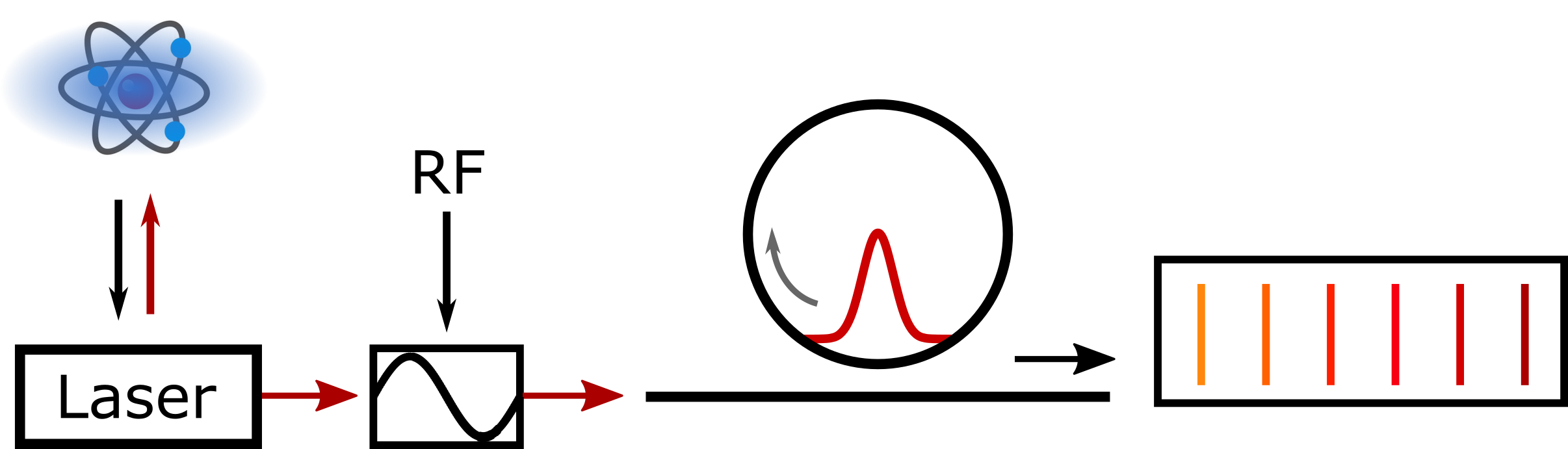}
    \caption{Schematic of a ring-resonator based photonic frequency comb for astronomy. The setup shown here is (simplified, not to scale) based on the astrocomb POCO at AIP~\cite{Boggio:2022,Madhav:2023}.}
    \label{fig:FC}
\end{figure}

Frequency combs --so called \textit{Astrocombs}-- can provide such a precise ruler, and different frequency comb technologies exist. One example are frequency combs based on ring resonators on a photonic chip~\cite{Obrzud:2019, Boggio:2022}, where the small ring structure (see Fig.~\ref{fig:Schem-structures} for an illustration) leads to a reference spectrum with a suitable line spacing (on the order of tens of GHz) for applications in astronomical spectrographs. Here, a pump laser is modulated (using radio frequencies, RF) and coupled into a WG, from where the light evanescently couples into a nearby ring waveguide, see Fig.~\ref{fig:FC}. Inside this resonator, an output of evenly spaced lines is formed that can be used as a reference spectrum. High precision of the reference spectrum can be achieved by stabilizing the pump laser to an optical atomic transition (e.g. using a rubidium gas cell) to avoid drifts of the center frequency and the RF modulation signal to an atomic RF clock to avoid changes in the output line separation by stabilizing the repetition frequency. Here, by utilizing the global positioning system (GPS) for external time keeping, long-term stability can be achieved. Astrocombs can thus provide stable reference spectra for spectrographs and are already being applied and verified~\cite{Madhav:2023,HARPS:2024,Probst:2020}. Ongoing research aims to, e.g. increase the flatness of the output spectrum~\cite{Wu:2024}, cover a wider wavelength range, and to create long-term stable, turn-key instruments. 

\subsection{Astrophotonics for Interferometry}
\label{astrophotonicsforinterferometry}
Optical and infrared interferometry~\cite{Monnier:2003,eisenhauer:23} --see bottom right of Fig.~\ref{fig:astrobeampath} (label~6) for an illustration-- is at the forefront of not only high angular resolution imaging~\cite{Ibrahim:2023, Martinez:2021} (enabling to identify details that are otherwise too small to resolve), but also high contrast imaging (enabling to detect low brightness object in the vicinity of a bright source) or a combination of both~\cite{LacourExo:2019}. Light from different telescopes (or sub-apertures) is interferometrically combined (within the coherence length of the light) and the visibility is extracted for different baselines (defined by the distance between the telescopes or sub-apertures). This is a measure of the coherence of the source and can be used to obtain information about the source's brightness distribution. How the telescope light is brought to interference, i.e. the beam combination, is a key aspect of interferometric instruments. In general, interferometric techniques are widely used in instrumentation, not only for scientific observations, but also for sensing, metrology, and stabilization. Not all of these are covered in this paper -- here, photonic solutions for three specific interferometric categories are briefly presented in the following sections:
\begin{itemize}
    \item aperture masking / pupil remapping interferometry in Sec.~\ref{AMI},
    \item long baseline interferometry (specifically amplitude interferometry) in Sec.~\ref{OLBIN},
    \item nulling interferometry in Sec.~\ref{NI}.
\end{itemize}

\subsubsection{Reformatters for pupil remapping interferometry}
\label{AMI}
Aperture masking interferometry~\cite{Sallum:2024} describes a technique in which a mask covers most of the telescope's aperture, letting light pass only through a set of non-redundant sub-apertures (redundant sets increase contributions from atmospheric noise~\cite{Tuthill:2010}). Light from these sub-apertures is then interferometrically combined. This technique removes atmospheric noise and has a narrower core of the point-spread function (PSF), avoiding the full telescope Airy disc near the center that makes the detection of close companions of a central star difficult. It enables high resolution images, even below the Rayleigh diffraction limit of the full telescope aperture~\cite{Lacour:2011}. As the majority of light is usually blocked by the mask, it is best suited for bright targets. This technique is even used on-board the James Webb Space Telescope~\cite{2016jdox} (JWST), which has an instrument with an aperture masking interferometer mode, NIRISS~\cite{2016jdox,Artigau:2014,Pope:2024,Cooper:2024}, for high angular resolution observation of companions (i.e. exoplanets) around stars.  

A variation of this technique is pupil remapping interferometry~\cite{Perrin:2006,Lacour:2006,Kotani:2009}, where the telescope pupil is divided into a (redundant) set of sub-pupils, which can in principle cover the whole telescope. Light from each of these pupils can be captured, remapped using photonics, and combined. This is where the confinement of light in photonic WGs and their ability to redistribute make a difference: using fibers or other (3D) waveguides allows for reformatting (see Fig.~\ref{fig:astrobeampath}, label 2) of the sub-pupils from a redundant to a non-redundant configuration for interferometry. This can increase the fraction of light used from the telescope, and thus the measurement sensitivity - particularly important for faint objects. Additional signal-to-noise improvements come from spatial filtering~\cite{Tuthill:2010, Perrin:2006} inherent to SM waveguides, which removes spatial phase fluctuations~\cite{Huby:2012} (speckles). 

Two advanced photonic based pupil remapping interferometry instruments are FIRST and Dragonfly (see also Ref.~\citenum{Tuthill:2010}):
\begin{itemize}
    \item FIRST~\cite{Huby:2012, Vievard:2020} (Fibered Imager foR a Single Telescope) operates in the visible wavelength range, using SM fibers for pupil remapping. FIRST has been tested on-sky and is now installed at the 8.2-m Subaru~\cite{Subaru} telescope (as part of the Subaru Coronagraphic Extreme Adaptive Optics (SCExAO) instrument~\cite{Vievard:2020}), where it can utilize $2 \times 9$ sub-apertures for high contrast imaging and spectroscopy~\cite{Huby:2024b}. As an upgrade (FIRST-v2/FIRST-PIC), a photonic chip is foreseen for beam combination of --currently-- five inputs~\cite{Barjot:2020, Kenchington:2024}, with different (lithographic~\cite{Barjot:2020, Kenchington:2024} and ULI~\cite{Martin:2024}) developments and their characterization still ongoing. This is particularly interesting as there are not many examples of astrophotonics in the visible wavelength range~\cite{Huby:2024}.
    \item Dragonfly\cite{Jovanovic:2012,Norris:14, Cvetojevic:2021} operates in the NIR, using laser-written SM waveguides in a glass substrate for pupil remapping. The photonics remapper was written in a boro-alumino-silicate glass substrate, with a foot print on the cm~scale and a thickness of 1.1 mm~\cite{Jovanovic:2012,Cvetojevic:2021}, and was optimized for the H-band. A ULI pupil remapper with eight WGs has been tested on-sky~\cite{Jovanovic:2012}, and a version of the chip has subsequently been interfaced with a lithographically fabricated, eight-input photonic beam combiner~\cite{Cvetojevic:2021}, to form a more integrated, hybrid instrument, which has been tested in the laboratory. The ULI-fabricated pupil remapper of Dragonfly was a precursor to the reformatting and beam combining photonics in GLINT~\cite{Lagadec:2021} (see Sec.~\ref{NI}).
\end{itemize}
In both cases --as for all interferometry instruments--, path length matching (here: for the reformatters) plays an important role~\cite{Tuthill:2010}, which is harder to obtain and maintain for fibers (e.g. by fiber polishing~\cite{Vievard:2020} and by adding a fiber delay line, respectively) than for laser-written WGs in monolithic glass blocks. On the other hand, for laser written WGs in glass substrate, there are different effects that have to be identified and avoided, e.g. bend-losses and stray light (which can be reduced through a "side-step" in the remapper~\cite{Norris:14}). Coupling into either fiber or laser-written reformatters can be done using micro-lens arrays, where the micro-lenses define the sub-pupils~\cite{Huby:2012}. 

Given the freely configurable 3D-WG writing capabilities of ULI, input reformatting and beam combination can be performed in the same photonic chip. This has been realized in different configurations for four~\cite{Nayak:21} and six~\cite{Dinkelaker:2023} input beam combiner devices, where input and/or output regions contain reformatters. Note that reformatters can not only be used for interferometry, but in all kinds of instrument to obtain a favorable configuration, e.g. to interface with v-groove fiber arrays or micro-lens arrays of a specific pitch, for use with spectrographs~\cite{Pike:2020} (e.g. Integral-Field Units (IFU))~\cite{Haffert:2020} and to better utilize detector space (in particular when light is also spectrally dispersed).

\subsubsection{Long baseline interferometry}
\label{OLBIN}

For long baseline interferometry, light from $N$ different telescopes in an array is routed to a beam combination instrument. Here, the $\frac{N\cdot(N-1)}{2}$ beam combinations can take place using classical free space optics~\cite{Brummelaar:2013} (e.g. mirrors, beam splitter plates or cubes) or photonics. 

\begin{figure}[h]
    \centering
    \includegraphics[width = 0.99 \linewidth]{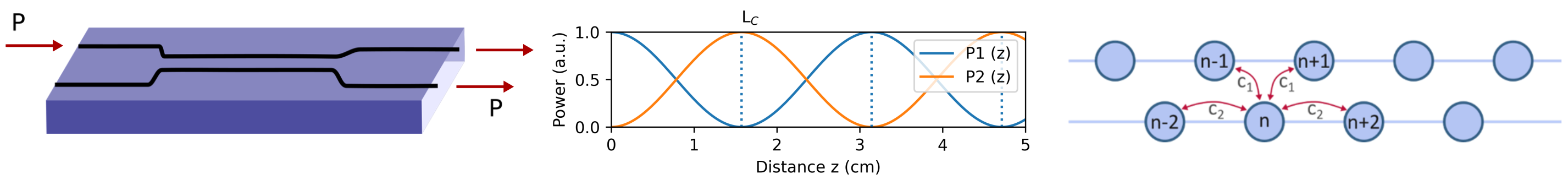}
    \caption{The foundation for photonic beam combiners is the directional coupler, where light in one WG can couple evanescently into a second WG nearby. Depending on the geometry of the interaction region, some or all of the light is transferred. Factors such as length of the interaction region and distance between WGs can be tuned to create a 50:50 beamsplitter, which means that when light is injected into both inputs, it can be evenly combined. (Left:) Schematic of a directional coupler, with an interaction region in the center. (Center:) Ideal optical power transfer curve in a (simplified) directional coupler from one WG to the other as a function of position $z$ along the WG. The coupling length $L_C$ denotes when 100\% of light is transferred. (Right:) This concept of evanescent coupling can be used not just for two but for many nearby WGs, even in 3D, where $c$ denotes the coupling strength (illustration based on Ref.~\citenum{Osellame:2012}).}
    \label{fig:DC}
\end{figure}

\begin{figure}[h]
    \centering
    \includegraphics[width = 0.75 \linewidth]{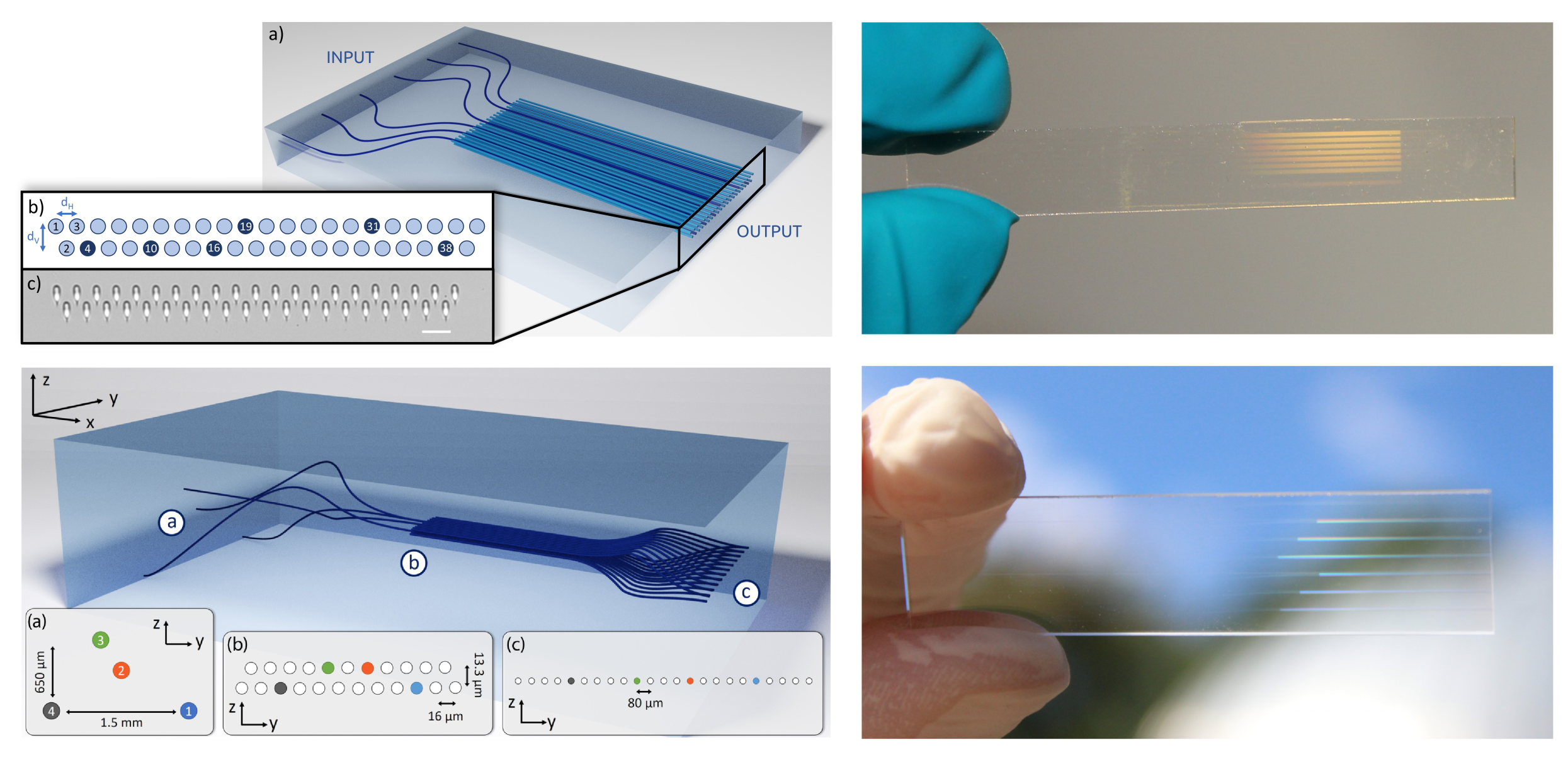}
    \caption{DBCs~\cite{Minardi:2010} are examples of 3D photonic chips. (Top left:) Schematic of a six-input DBC for the J-band (reprinted with permission from Ref.~\citenum{Dinkelaker:2023} \copyright~Optica Publishing Group). The beam combiner array consists of 41~WGs; the arrangement in two layers can be seen in the insets b) and c), an input reformatting region allows interfacing with fiber arrays. (Top right:) Photograph of a chip containing multiple DBC arrays. The device has been fabricated by Politecnico di Milano in borosilicate glass using ULI. (Bottom left:) Illustration of the four-input DBC for the H-band (image from Ref.~\citenum{Nayak:21}). The same chip contains reformatters at the input and output regions (insets a and c). (Bottom right:) Photograph of a photonic chip with several four-input DBCs, chip fabricated by Politecnico di Milano. The photonic structures become visible in the photograph as a each of the WG arrays acts as a little grating for the light. The input reformatters are faintly visible coming on the left side of the chip. Six WG arrays of different lengths can be seen on the right.  }
    \label{fig:DBC}
\end{figure}

While free-space optics have the advantage of being easily accessible and have a high transmission, individual components are typically on the cm-scale, the setup quickly fills up a large part of the optical table and can become very complex - especially for larger number of telescope inputs~\cite{Berger:2001}. In addition, vibrations and temperature changes might affect the individual paths differently, increasing noise. Photonic components can address several of these aspects: as for pupil remapping interferometry, spatially filtered light (through SM waveguides) will lead to increased precision. The compact design can drastically decrease the required space of the setup, and fiber interfacing means that different sub-systems can be decoupled, which facilitates changes and upgrades within one subsystem without disturbing the other (e.g. replacing the output optics without changing any optics that come before the fiber coupling). There are different photonic technologies for beam combination, which in some cases have already been implemented in science instruments, either utilizing photonic chips or fibers. Beam combiners in photonic chips are based on directional couplers, where light is coupled between close-by WGs (inter-WG distance on the order of ten $\mu$m) through evanescent coupling (see Fig.~\ref{fig:DC}). We can distinguish the following photonic beam combiner types:

\begin{itemize}
    \item Photonic beam combination in \textbf{integrated optics chips}: 
    \begin{itemize}
        \item \textbf{2D} (planar) beam combiner chip. The most well-known example of this time is probably the beam combiner in the GRAVITY~\cite{Gravity} instrument at VLTI. The instrument operates in the K-band ($2 - 2.4~\mu$m) and demonstrates the capabilities of its photonic, four-input, planar beam combiner chip~\cite{Benisty:2009}, which instantaneously samples a fringe for each pairwise beam combination (ABCD sampling). The beam combiner is interfaced with a fiber array and cryo-cooled (at 200~K). Earlier developments~\cite{Kern:1997, Malbet:1999} resulted in the first prototypes of planar integrated optics chips with two and three inputs (H-band) that were tested at the telescope, namely the Infrared Optical Telescope Array (IOTA)~\cite{Berger:2001,Berger:2003}, and implementation of an integrated optics chip in the PIONIER instrument at VLTI~\cite{Bouquin:2011}. Currently, a planar six-input chip (SPICA-FT)~\cite{Pannetier:2022} for the H-band is under commission as a fringe tracker for SPICA~\cite{Mourard:2024}.
        \item \textbf{3D} beam combiner chip. Examples of beam combiners making use of the 3D capabilities of ULI are discrete beam combiners (DBC)~\cite{Minardi:2010}, which evanescently couple light from the input WGs through an array of WGs, avoiding any WG crossings. The WGs are distributed over multiple stacked (and potentially horizontally shifted) layers, and their 3D structures make them potential candidates for a larger number of telescopes. DBCs have been developed for different wavelengths and number of inputs: three inputs for visible~\cite{Saviauk:13} in fused silica, six inputs for J-band~\cite{Dinkelaker:2023} in borosilicate glass, four inputs for H-band~\cite{Nayak:21} in borosilicate glass, four inputs for L-band~\cite{Diener:17} in Gallium Lanthanum Sulfide (GLS) glass. The chips have footprints on the cm-scale, with a thicknesses on the order of 1~mm. Several of these devices have reformatting regions inside the same chip at the input (path-length matched) and output. A recent analysis~\cite{Dinkelaker:2023} of the six-input DBC fabricated using ULI showed a throughput of around $56\%$ in the J-band, and an instrumental contrast of $89\% \pm 7\%$ (comparable to the lithographic SPICA-FT chip~\cite{Pannetier:2022}). Two examples of DBCs~\cite{Dinkelaker:2023, Nayak:21} are shown in Fig.~\ref{fig:DBC}. A 3D chip for reformatting and nulling interferometry is the ULI-fabricated GLINT beam combiner chip, see Sec.~\ref{NI}.

    \end{itemize}
\end{itemize}


\begin{itemize}
        \item Photonic beam combination using \textbf{fibers}: 
    \begin{itemize}
        \item \textbf{Fiber-coupling and reformatting} of the telescope light to prepare a fiber array, with subsequent multi-axis beam combination in the image-plane. The joint instrument MIRC-X / MYSTIC (H-band and K-band, respectively) at CHARA utilizes this technique, where optical fibers prepare the light in a non-redundant configuration inside a v-groove array for image-plane beam combination of six telescope beams~\cite{Monnier:2018,Anugu:2020}. (For MYSTIC, a four-input GRAVITY-like chip is available as well.) Parts of the MYSTIC instrument are cryo-cooled, including the beam combiner, where the fibers enter the cold optics system with a fiber feedthrough~\cite{Monnier:2018}.
        \item \textbf{In-fiber} beam combination. The FLUOR instrument~\cite{Foresto:1997,Foresto:2003} (first at Kitt Peak, then IOTA and CHARA) and its upgrade JouFLU~\cite{Scott:2013} used a fluoride fiber beam combiner for two telescope inputs in the K-band. FLUOR also demonstrated the improvement in visibility precision due to spatial filtering in SM fibers~\cite{Foresto:1997,Foresto:1997b,Foresto:1997c}. The fiber beam combiner has degraded over time. In its place, a chip beam combiner has recently been installed into the setup (with additional modifications to the setup, FLUOR/JouFLU is now CHARIOT, see below). 
    \end{itemize}

\end{itemize}

\paragraph{A two-input ULI beam combiner for the K-band for CHARIOT at CHARA:} Recently, as part of the NAIR-2~/ APREXIS project, the FLUOR~/ JouFLU instrument at CHARA has been substantially modified and is now known as CHARIOT~\cite{Mayer:2024}: the degraded fiber beam combiner (MONA) was replaced by a fiber-connectorized, ULI-fabricated, two-input beam combiner for the K-band (fabricated by Heriot-Watt University (HWU))~\cite{Benoit:21,Siliprandi:2024}. In addition, a C-RED One camera was installed, and the beam path layout on the optical table was re-designed. Currently, the setup consists of two length-matched beam paths (including motorized mirrors and translation stages), and a design that already accounts for the addition of two more beams, as well as a fiber coupling system and an output optics stage, see also Ref.~\citenum{Mayer:2024}. In future, CHARIOT is planned to enable four-telescope interferometry in the J-,~H-,~and K-band. The CHARIOT setup has two main goals:
\begin{enumerate}
    \item Perform on-sky tests and verification of the ULI-fabricated, two-input beam combiner for the K-band.
    \item Create the CHARIOT optical table as a bench that allows easy integration (ideally plug-and-play using fibers) for on-sky tests of (nulling) beam combiners and other astrophotonic components for the interferometry community at CHARA~\cite{Mayer:2024}.
\end{enumerate}

\begin{figure}[b]
    \centering
    \includegraphics[width = 0.9 \linewidth]{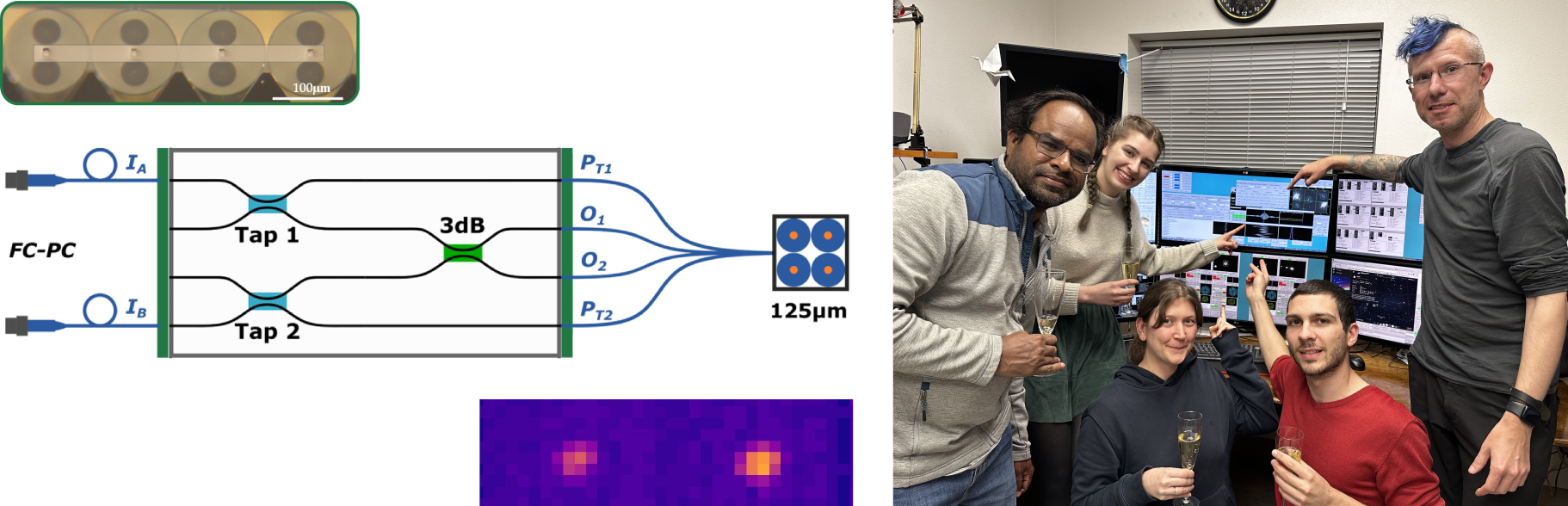}
    \caption{(Left:) Schematic of the ULI-fabricated, fiber-connectorized, two-input beam combiner, image taken from Ref.~\citenum{Siliprandi:2024} and modified; the upper inset shows an image of the fibers in the v-groove overlayed with the chip WGs~\cite{Siliprandi:2024}. The bottom inset shows an image of the two interferometric outputs with a calibration source in CHARIOT. (Right:) Celebrating first on-sky fringes in the K-band at CHARA with such an ULI beam combiner in the CHARIOT setup~\cite{Mayer:2024}. From left to right: Narsireddy~Anugu (GSU/CHARA), Alyssa~V.~Mayer (AIP), Aline~N.~Dinkelaker (AIP), K\'evin~Barjot (University of Cologne), and Nicholas~J.~Scott (GSU/CHARA), image credit: Gail~Schaefer (GSU/CHARA).}
    \label{fig:fringes}
\end{figure}

The first goal of CHARIOT has been to perform on-sky tests of the first ULI-fabricated beam combiner for the K-band. To extend the accessible wavelength range for photonic components, e.g. achieve high throughput towards the longer wavelength end of the K-band, new materials and fabrication methods have to be explored. To make best use of the K-band, an achromatic directional coupler with $50:50$ splitting and low losses has to be fabricated. Here, Infrasil\textsuperscript{\textregistered} glass (IG) is a suitable substrate material, with flat transmission of around 90\% from $200$~nm to~$> 3000$~nm and low water~/~OH content (which means that it does not have absorption losses around e.g. $2.2~\mu$m that can be found in fused silica). Following a parameter study, an asymmetric two-input coupler with photometric taps was fabricated with ULI by HWU, with an refractive index change between the IG substrate and the WGs of around $8 \times 10^{-3}$. A schematic~\cite{Siliprandi:2024} is shown in Fig.~\ref{fig:fringes}; the fabrication and characterization details can be found in Refs.~\citenum{Benoit:21,Siliprandi:2024}. Selected devices were then aligned and glued to polarization maintaining NUFERN~PM1950 fibers in a v-groove array (by OzOptics) with connectorized, length matched pairs (length matching by FOC) at the input. At the output, the four output fibers from the beam combiner chip were arranged into a tight $2 \times 2$ array for imaging, using a tapered surface channel that was fabricated in a fused silica wafer using ULAE (see Sec.~\ref{fab}). Characterization of the beam combiner revealed an achromatic $50:50$ splitter with a throughput of around $77\%$, and an interferometric contrast of $80 - 90 \%$~\cite{Benoit:21,Siliprandi:2024}. One of the fiber-connectorized chips was integrated into the CHARIOT setup at CHARA, light from laboratory calibration sources was coupled into the system, and an output optics system was aligned. Following beam path length matching and laboratory characterization, first on-sky fringes were achieved for this ULI-fabricated, fiber-connectorized, two-input beam combiner for the K-band (see Fig.~\ref{fig:fringes}). The data analysis is currently in progress.

\subsubsection{Nulling interferometry}
\label{NI}
A specific subcategory of interferometry (either long baseline or aperture masking interferometry) is nulling~\cite{Bracewell:1978}, where beam combination is phase shifted in such a way that on-axis star light is destructively interfered and off-axis light is transmitted. This technique aims at high contrast imaging of faint objects, i.e. exoplanets, close to the host star. For exoplanet detection, the required contrast depends on the type of planet --from $10^{-4}$ down to $10^{-10}$-- and the resolution depends on the planet's orbit, see e.g. Ref.~\citenum{Norris:2019b,Martinod:2021}. Nulling interferometers with free-space optics exist, but phase errors reduce the achievable star light suppression. With photonic nulling beam combiners that incorporate the design phase shift in their WGs, spatial filtering can decrease phase fluctuations and improve suppression. Different photonic nullers are currently being developed:

\begin{itemize}
    \item The GLINT (Guided Light Interferometric Nulling Technology)~\cite{Norris:2019b,Martinod:2021} instrument is integrated into the SCExAO system at the Subaru telescope (with its pathfinder instrument \textit{GLINT South} to test prototypes at the 3.9-m Anglo-Australian Telescope)~\cite{Lagadec:2021}. GLINT employs ULI-fabricated, 3D photonics for reformatting of the injection sites and beam combination, building on the technology developments of Dragonfly~\cite{Lagadec:2021} (see Sec.~\ref{AMI}). GLINT combines --and nulls- light from four inputs simultaneously in the H-band~\cite{Martinod:2021}. A two-input version of the chip was previously tested~\cite{Norris:2019b}, incorporating a side-step structure to avoid scattered light, with a relative $\pi/2$~rad delay at the directional coupler, and four outputs: two photometric outputs, one \textit{dark} (null), and one \textit{bright} (anti-null) output~\cite{Norris:2019b}. The nulling combiner was written in a boro-alumino-silicate glass with $41~\textrm{mm}\times  10~\textrm{mm} \times 0.7~\textrm{mm}$, with an internal transmission of $85~\%$. The four-input chip~\cite{Martinod:2021} has dimensions of $ 75~\textrm{mm}\times 3.5~\textrm{mm} \times 2$~mm and has 16 outputs (six of which are the null outputs), which are fiber coupled to feed a spectrograph. Both chips have been tested on-sky where they measured stellar diameters below the diffraction limit of the telescope. To increase achromaticity and increase sensitivity, a 3D tricoupler~\cite{Martinod:2021b} has been developed for application in GLINT. See also Refs.\citenum{Spalding:2024,Arcadi:2024,Arcadi:2024b,Douglass:2024} for updates. 
    \item The Palomar Fiber Nuller~\cite{mennesson2011high} is an instrument for the K-band at the Palomar Hale telescope that utilizes a fiber for beam combination: light from two sub-apertures and a relative $\pi$~rad phase shift is injected into one SM fiber (i.e. multi-axial beam combination). The telescope pupil is then rotated to obtain measurements of regions around stars~\cite{kuhn2015exploring}. 
    \item NOTT (Nulling Observations of exoplaneTs and dusT, formerly Hi-5) is planned as a future high contrast nulling beam combiner~\cite{Sanny:2022, Laugier:2023, Garreau:2024,Defrere:2024} that will be part of the ASGARD suite for the VLTI~\cite{Martinod:2023}. It is designed to operate in the L' band ($3.5 - 4.0~\mu$m), i.e. in MIR, for four telescope inputs, and a photonic chip~\cite{Sanny:2022} is planned to be used as beam combiner. Prototype devices have been manufactured using ULI in GLS glass and characterization has been performed in the laboratory. The design includes a side-step to avoid stray light, as well as a Double Bracewell sequence of directional couplers, see Ref.~\citenum{Sanny:2022} for details and Refs.~\citenum{Sanny:2024,Bonduelle:2024} for updates on photonics development.
    \item LIFE is an initiative working towards a space-based nulling interferometer in the wavelength range $4 - 18.5~\mu$m (MIR) for exoplanet detection~\cite{LIFE, Quanz:2022, Glauser:2024}. Photonic technologies are good candidates for space missions, and the options for components~\cite{Ireland:2024} and mission design are currently being developed and evaluated, e.g. a five-telescope kernel nulling interferometer~\cite{Hansen:2023}. Technology development for other instruments, e.g. for NOTT, will benefit the realization of LIFE.  
\end{itemize}

\section{DISCUSSION AND PERSPECTIVES}
\begin{figure}
    \centering
    \includegraphics[width=0.85\linewidth]{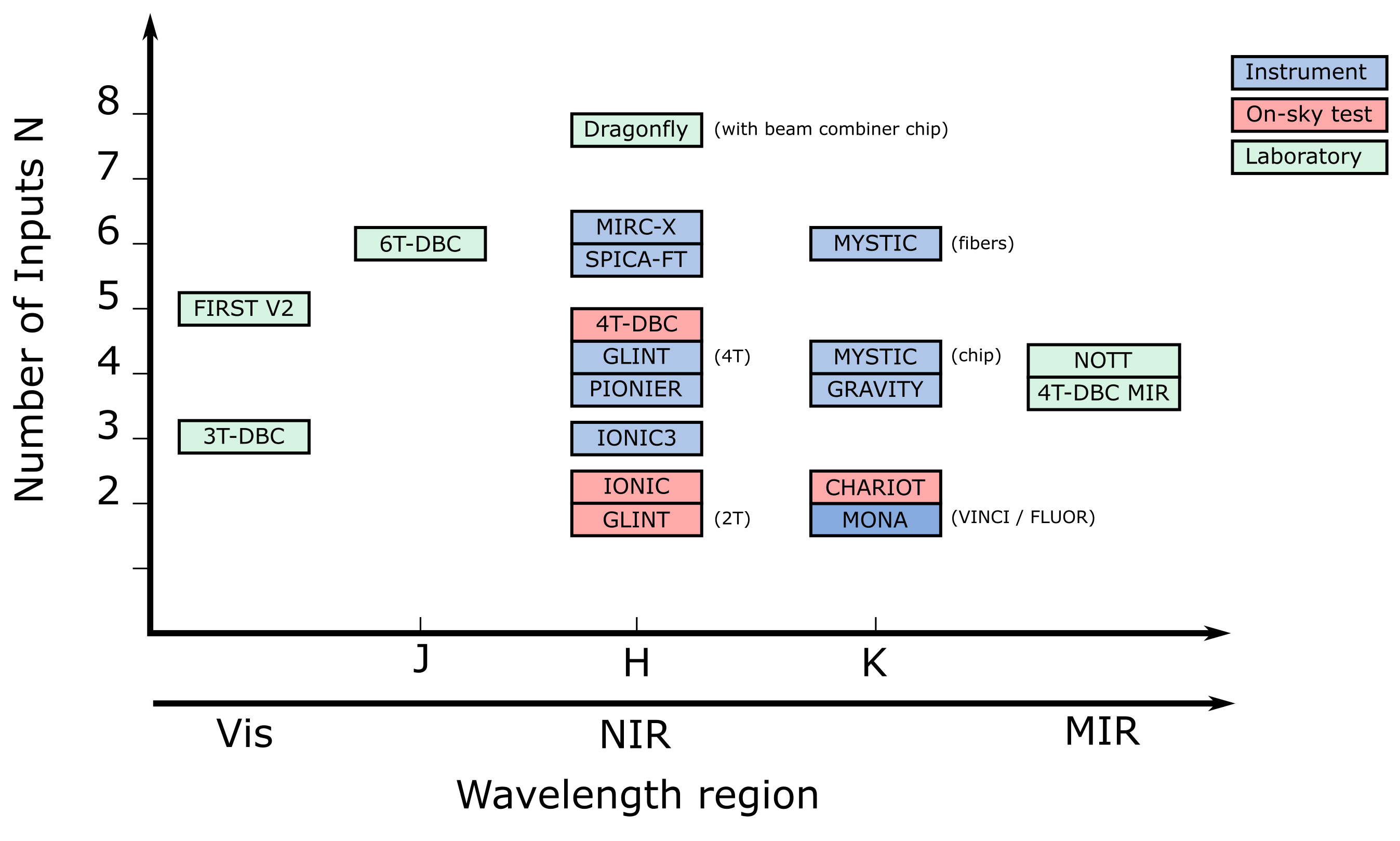}
    \caption{Overview of the photonic beam combiner landscape, based on Ref.~\citenum{Labadie:2022}. See text for descriptions and details on individual beam combiner technologies.}
    \label{fig:BC}
\end{figure}

For applications in interferometry, the photonic beam combiner landscape is getting more diverse with respect to accessible wavelengths and number of inputs. A (probably incomplete) overview is shown in Fig.~\ref{fig:BC} (based on figures in Ref.~\citenum{Labadie:2022}, where more details can be found), but the graph has to be populated along both axes if future instrument requirements are to be fulfilled (e.g. efficient devices at longer wavelengths for the LIFE space mission). 

When it comes to integration into an instrument at a telescope, there are more things to consider than the performance of the bare photonic component itself, e.g. the stability of light coupling, ease of interfacing or data extraction. On-sky tests play a crucial role, not only to verify that a component works with star light. The situation of using a component in an instrument at the telescope can be vastly different to laboratory testing, with low photon levels, temperature variations (e.g. night time vs. day time) that may lead to decoupling~\cite{Nayak:21}, atmospheric turbulence, small available space, limited time window for observation, etc. In addition to technological verification, on-sky tests (and their preparations) thus provide important training opportunities for researchers to consider next generation designs of components, software, and packaging, e.g. selecting common fiber interfaces, expecting realistic signal and noise levels (as compared to the ideal lab situation with calibration sources), enabling fast access to data and data reduction to allow instantaneous optimization if possible. The CHARIOT testbed~\cite{Mayer:2024} at CHARA is one example, how this hands-on experience can be facilitated. This is in line with other efforts to increase training opportunities and to utilize resources efficiently: following the \textit{2023 Astrophotonics Roadmap}~\cite{Jovanovic:2023} as well as a virtual workshop in January 2024, the 2024 SPIE Astronomical Telescopes~$+$~Instrumentation conference included an \textit{Astrophotonics Community Workshop} (organized by Kyler Kuehn, Lowell Observatory). Here, participants had the opportunity to meet in person and discuss topics such as access to community-wide resources, plans as to how technical training in astrophotonics can be enhanced (especially for early career researchers), as well as technical goals and challenges within the field.

\paragraph{Challenges and Opportunities:}
There are several topics that have not been discussed here, e.g. active photonics, the challenge of micro-integration of detectors, the potential of technology transfer (e.g. quantum technology or medical instruments) or extended applications of components beyond what was initially thought (such as the use of PLs as wavefront sensors~\cite{Norris:2024}). There are also a range of general goals and challenges that will require continuous improvements and engineering solutions, e.g.:
\begin{itemize}
    \item stable and efficient free space to WG coupling / fiber to chip coupling,
    \item cryogenic operation of devices and instruments,
    \item packaging design for turn-key operation and user friendly, stable operation,
    \item increased throughput,
    \item increased broadband operation,
    \item deterministic and repeatable fabrication.
\end{itemize}

Many of these topics can be found in more detail in the \textit{2023 Astrophotonics Roadmap}~\cite{Jovanovic:2023}. This paper aimed to provide some of the many examples of astrophotonics, as well as present some of the underlying aspects of the two fields, astronomical instrumentation and photonics. There are different reviews and collections on the topic for more detail and a wider range of examples (see references in Sec.~\ref{sec:intro}). Astrophotonics will continue to evolve, hopefully creating more integrated instruments and technical solutions to push sensitivity and resolution limits and support the ambitious goals in astronomy and astrophysics. 

\section*{ACKNOWLEDGEMENTS}
 
This work was supported by the Deutsche Forschungsgemeinschaft (DFG) under grant number 506421303 (\mbox{NAIR-2~APREXIS}), which is a project by the University of Cologne, Leibniz Institute for Astrophysics Potsdam (AIP), and Durham University, in close collaboration with CHARA and Heriot-Watt University. The research leading to some of the results presented here has received funding from the European Union’s Horizon 2020 research and innovation programme under Grant Agreement 101004719 (ORP). Aline would also like to thank Kalaga Madhav, Alyssa V. Mayer and the whole Astrophotonics group at AIP for input and discussions, as well as Lucas Labadie and his group at the University of Cologne, Aurélien Benoît and Robert R. Thomson at HWU, and the staff at CHARA. 

\bibliography{report} 
\bibliographystyle{IEEEtran} 

\end{document}